# Evaluating Digital Inclusiveness of Digital Agri-Food Tools Using Large Language Models: A Comparative Analysis Between Human and AI-Based Evaluations

Pewinya, Githma

Colombo,

Sri Lanka

githmam00@gmail.com

Iglésias Martins, Carolina

International Water Management Institute

C.Martins@cgiar.org

Garcia, Mariangel Andarcia

International Water Management Institute

M.GarciaAndarcia@cgiar.org

# Evaluating Digital Inclusiveness of Digital Agri-Food Tools Using Large Language Models: A Comparative Analysis Between Human and AI-Based Evaluations

## Abstract

Ensuring digital inclusiveness is a critical priority in agri-food systems, particularly in the Global South, where digital divides persist. The Multidimensional Digital Inclusiveness Index (MDII) offers a comprehensive, human-led framework to assess how inclusive digital agricultural tools (agritools) are. However, the current evaluation process is resource-intensive, often requiring months to complete. This study explores whether large language models (LLMs) can support a rapid, AI-enabled assessment of digital inclusiveness, complementing the MDII's existing workflow. Using a comparative analysis, the research benchmarks the performance of four LLMs (Grok, Gemini, GPT-4o, and GPT-5) against prior expert-led evaluations. The study investigates model alignment with human scores, sensitivity to temperature settings, and potential sources of bias. Findings suggest that LLMs can generate evaluative outputs that approximate expert judgment in some dimensions, though reliability varies across models and contexts. This exploratory work provides early evidence for the integration of GenAI into inclusive digital development monitoring, with implications for scaling evaluations in time-sensitive or resource-constrained environments.

## 1. Introduction

In an increasingly interconnected world, digital technologies play a central role in several industries, such as agriculture, health, education, and governance. While digital technologies allow increased access to information, services, and economic opportunities, many of these technologies can widen the digital gap by reinforcing inequalities rather than reducing them (Djatmiko et al., 2025). Due to the inherent complexity of socio-technological factors that enable or hinder equitable adoption, ensuring that no group is digitally excluded remains a critical research and policy priority. Therefore, digital inclusiveness has become a critical determinant for equitable access to technological innovations and potential benefits. To support a more inclusive digital ecosystem in the Global South, the Multidimensional Digital Inclusiveness Index was developed to provide a systematic framework for evaluating the inclusiveness of digital tools in the agricultural sector (Martins et al., 2024; Opola et al., 2025). In its current form, the MDII expands on traditional digital evaluation frameworks by considering dimensions beyond accessibility and expanding on enabling environment factors. It considers seven dimensions (Beneficial Impact; Usage Effectiveness; Ethical and Responsible Innovation; Co-Creation and Governance; Risks and Harms; and Supportive Ecosystem) across three mega-groups (Innovation Usage;

Stakeholder Relationships and Beneficial Impact). These dimensions are further operationalized into 24 subdimensions, which are evaluated through data drawn from 90 indicators, each defined by targeted evaluation questions, as depicted in Figure 1. As an index, the MDII's purpose is to allow harmonization of the evaluation of digital agritools. This harmonization leads to the ability for tools to be compared against a quantitative metrics, supporting tailored recommendations and allowing a comprehensive dataset of tools to be built over time.

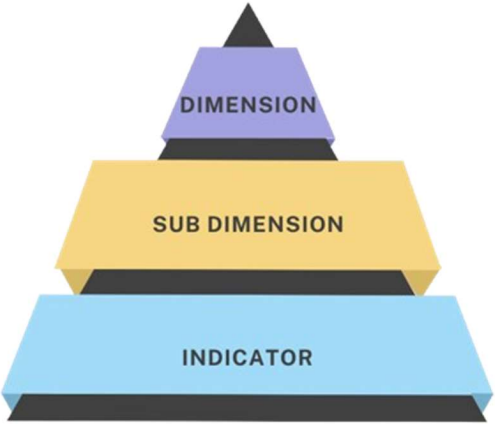

*Figure 1: Hierarchy of MDII components. Source: Adapted from (Martins, et al., 2024)*

To capture inclusiveness comprehensively within digital agritools, the MDII integrates the perspectives of four stakeholder groups (Figure 2): (i) tool innovators or developers (responsible for creating digital tools); (ii) domain expert evaluators (professionals with expertise in assessing performance and inclusiveness); (iii) direct user evaluators (individuals who actively engage with the tool); and (iv) downstream beneficiaries (individuals that don't use the tool are impacted by its outputs). By combining these diverse perspectives, it becomes possible to generate a system-and-evidence-based assessment of digital tools, enabling innovators to identify inclusiveness gaps, increase equity, and accelerate the implementation of inclusive digital ecosystems.

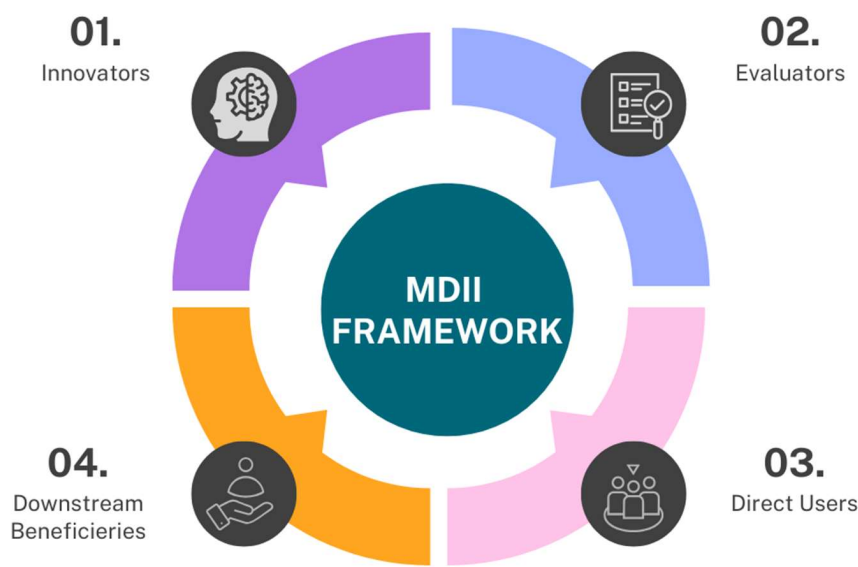

*Figure 2: Stakeholder groups. Source: Adapted from (Martins, et al., 2024)*

Currently, the MDII's evaluation workflow relies entirely on data provided by humans. While this full evaluation allows for a comprehensive inclusiveness assessment, it is resource-intensive. Field application often requires field enumerators, travel logistics, and sustained engagement, which results in high operational costs. Moreover, the length of the evaluation process can take around 6 to 8 months from data collection to score delivery. Accounting for these constrains, the requirement from innovators for faster feedback loops, and recent advances in generative AI (GenAI), a new form of AI-enabled rapid assessment for MDII was explored as a complementary approach to the existing human-led workflow. This exploration is grounded in the following working assumption:

> ***Assumption 1: GenAI and current LLMs seem to be mature enough to support evaluation of digital inclusiveness due to their capability of being fine-tuned to their task.***

Traditionally, within the MDII framework, tools are evaluated by domain experts who act as impartial third parties, expected to provide a critical and informed perspective aligned with their field of expertise. These experts review qualitative responses submitted by innovators and assign scores across multiple indicators using structured Likert scales. While expert-led evaluations contribute valuable contextual understanding, they face key limitations: scoring inconsistencies, subjective interpretation, and reliance on the availability and responsiveness of evaluators. Furthermore, the data collection process is resource-intensive, logistically demanding, and subject to delays, especially when covering geographically dispersed tools and multilingual contexts. Additional constraints such as limited internet access or competing time commitments may further affect the completeness and comparability of assessments. To address these challenges, scalable alternatives that complement human judgment while preserving analytical rigor are needed. In this context, LLMs offer the potential to automate portions of the evaluation process by processing large volumes of qualitative data, automatically applying scoring prompts, and generating consistent, criteria-based outputs (Dunivin, 2025). This rationale led to a second working assumption:

> ***Assumption 2: Integrating LLMs into the MDII assessment process could enhance efficiency, comparability and turnaround times for in-development tools.***

These two assumptions form the basis of this exploratory research. While the use of GenAI in evaluation remains experimental and results should be interpreted cautiously due to potential inaccuracies or misalignment, recurring feedback from innovators has highlighted the need for faster feedback cycles during development. From an ethical standpoint, this motivated the investigation of AI-enabled evaluation methods that might offer earlier-stage insights, supporting more inclusive and adaptive digital tool design prior to large-scale deployment.

## 2. Research Objectives

This study presents a comparative analysis between human and AI evaluations to assess the potential of LLMs in evaluating digital inclusiveness within agritools. By benchmarking AI-generated scores against expert judgments, the study aims to examine the degree of alignment, identify divergences or biases, and generate insights into the potential of LLMs to contribute to digital development monitoring at scale. The primary objectives of this study are to evaluate the potential of LLMs to assess digital inclusiveness within the MDII framework. More specifically, the study seeks to:

(i) Evaluate the performance and reliability of different LLMs (Grok, Gemini, GPT-4o, and GPT-5) in generating MDII scores across multiple domains;
(ii) Understand the influence of temperature sensitivity parameters on the stability of model-generated evaluation;
(iii) Compare AI-generated outputs with previous scores obtained from former domain expert evaluations and observe either alignment or divergence;
(iv) Identify strengths and limitations of AI-enabled evaluations in digital inclusiveness of agritools.

These objectives led to the formulation of the following research questions, grounded in the earlier assumptions:

> ***RQ 1**: Can GenAI LLMs evaluate digital inclusiveness in a manner comparable to human evaluators supporting a "rapid assessment" format?*

*RQ 2: Are GenAI evaluations subject to similar biases as human-led assessments?*
*RQ 3: How do different LLMs architectures behave when applied to the MDII rapid assessment process?*

This study adopts an exploratory approach that focuses on whether an AI-enabled scoring algorithm can replicate or approximate human evaluations within the MDII framework. The benchmark for comparison remains the full human-led MDII evaluation, which is comprehensive and grounded in human judgment. In contrast, the AI-enabled assessments evaluated in this study simulate partial evaluations, focusing on domain-specific indicators only.

## 3. Literature Review

### 3.1 The role of digital inclusiveness in agri-food tools

Digital technologies have promised to transform agri-food systems in ways that contribute to more inclusive, economically viable, and environmentally sustainable futures. However, without critical attention to how these technologies are introduced, developed, and rolled out, they risk reinforcing existing inequalities rather than addressing them. Digital transformation, when not guided by inclusive policies and regulatory frameworks, can exacerbate economic disparities rather than alleviate them (Adams et al., 2025). Specifically for the agri-food sector, a growing body of literature highlight that inclusive innovation frameworks often emphasize farmer participation but fall short in operationalizing the components for deeper inclusivity in a way that considers both the quality of engagement, and relational dynamics with smallholder farmers (Ndege et al., 2024). Consequently, even well-intentioned digital innovations can selectively include or exclude farmers depending on how usability, accessibility, and relevance are addressed (Ndege et al., 2024). The introduction of the concepts of "inclusive innovation *for*" and "inclusive innovation *with*" are presented as critical distinctions for understanding and improving inclusivity in practice. This perspective emphasizes the importance of advancing from superficial engagement to innovation processes that empower smallholder farmers and prioritize their agency in digital agri-food systems. Additionally, digital agriculture has been widely seen as a transformative force for global agri-food systems, however its adoption remains uneven across regions. In the Middle East and North Africa (MENA) region, digital technologies are still in early stages of implementation with adoption being mainly concentrated in high-value agricultural sectors and export-driven markets (Bahn et al., 2021). Evidence suggests that current efforts are primarily motivated by economic sustainability with limited focus on social or environmental inclusiveness (Bahn et al., 2021). This reinforces the need for policies that not only promote technological innovation but also ensure equitable access, transparency, data protection, and livelihood safeguards.

## 3.2 How can LLMs support digital inclusiveness

The increasing use of LLMs in human-centered domains has raised critical questions about their alignment with real human behavior, especially in decision-making contexts (Wolmetz et al., 2025). By using a novel Turing Representational Similarity Analysis (RSA), the author studied the feasibility of using LLMs as behavioral proxies in controlled experiments by comparing their internal representations and behavioral responses to those of humans. Findings suggest that while LLMs exhibit some alignment with human reasoning, their behavior diverges significantly in cases that involve individual personality traits and social dilemmas. These findings emphasize the importance of understanding the capabilities and the limitations of LLMs in evaluative roles. For AI tools deployed in agricultural settings, especially those related to smallholder farmers, the nuances of human-like behavior become critical for ensuring a context-sensitive decision support that is simultaneous ethical and interpretable.

With the fast-growing integration of AI in (digital) agriculture, ethical implications surround its use emerge as a major concern. AI-related risks can be categorized into six areas: fairness, transparency, accountability, sustainability, privacy, and robustness (Dara, 2022). Poorly designed AI systems may unintentionally harm farmers, animals, or ecosystems if not carefully managed (Dara et al., 2022). Examples relate to invasion of farmers' privacy, cause unintended consequences in robotic farming due to cybersecurity risks or design flaws (Dara et al., 2022). To mitigate these issues, agricultural technology providers and policymakers should focus on reducing bias and ensuring reliability in AI performance. Therefore, following the same principle within the MDII framework, an LLM application also requires a responsible and ethical approach in promoting a more equitable and trustworthy digital transformation of agriculture.

Apart from the application of AI-enabled evaluations in agricultural systems, the healthcare sector is considered a domain where AI-enabled tools can serve as effective additions in clinical practice by supporting clinical decision-making through evaluation and diagnostic accuracy. Although there are high stakes for the precision and reliability of health evaluation systems, Krakowski et al. (2024) found evidence on how deep learning-based AI assistance can bring significant improvements in clinician performance in skin cancer diagnosis. Across ten eligible studies, the author observed that pooled sensitivity and specificity for clinicians raised from 74.8% (without AI assistance) up to 86.1% (with AI assistance). Although studies evaluating how AI-enabled tools can support diagnosis accuracy were conducted in experimental settings, evidence suggests that integrating AI into routine clinical practice may enhance outcomes, if systems are rigorously validated and carefully implemented (Alowais et al., 2023). By considering that evaluating digital agritools under the MDII framework also requires comparable rigor to promote systemic digital inclusiveness, reliable and accurate evaluation methods are mandatory. Although inaccuracy in MDII evaluations can be considered less serious from the healthcare sector, in which errors in evaluation and decision-making can have profound devastating outcomes (e.g., as in clinical trials and assessments), the MDII (being by design an evaluation tool) has also the responsibility of precise and accurate scoring as that score can impact decisions in made in tool development.

### 3.3 Reflecting on AI and human evaluation bias

Mitigating bias in AI evaluation processes requires the detection of algorithmic shortcomings and inclusion of human expertise throughout the AI lifecycle (Weidinger et al., 2022). To address this need, Harfouche et al. (2023) proposed a human-centric AI framework to mitigate AI biases. The authors implemented the framework in two design science research (DSR) projects and verifies that the framework supports the integration of human knowledge in both the design and training of AI. These projects operationalized the framework through two complementary stages: first, by embedding a human-in-the-loop design process, and second, by creating a usage architecture that integrates human and AI collaboration. This approach enabled the identification and mitigation of biases during both the design and deployment phases of AI systems (Harfouche et al., 2023). The human-in-the-loop portion informed design phase, in which domain experts contributes to data curation, feature selection, and model validation. This was then followed by an augmented intelligence phase that enabled continuous human AI interaction in decision-making. In sum, both stages human knowledge to act as a corrective mechanism against algorithmic issues such as sample selection bias, out-group homogeneity bias, correlation fallacies, and apophenia. On the other hand, AI outputs were able counterbalance human cognitive errors. The study illustrates how varying degrees of human AI collaboration (from sequential decision-making to fully aggregated decision processes) enhances the effectiveness of bias mitigation measures. In the context of agrisystems, the same rationale of embedding domain-specific human expertise within AI-assisted evaluation systems could also be applied. Such integration might ensure that inclusiveness metrics are not only computationally robust but also contextually and socially attuned, reducing both algorithmic and evaluator-induced bias in the assessment of digital tools.

In addition, recent work on human-AI decision-making highlights that the effectiveness of AI systems is not only determined by their predictive accuracy but also by how users reconcile their own intuition with AI outputs. Work developed by Chen (2023) examined how decision-makers interpret and respond to AI-generated predictions when presented with different types of explanations. The study revealed three distinct intuition-driven strategies that participants used to override AI recommendations: (a) drawing on a strong intuitive sense of the correct outcome based on prior experience or domain knowledge; (b) engaging in feature-based reasoning by evaluating the relevance and weight of specific input features; and (c) identifying signs of prediction unreliability, such as inconsistencies or implausible outputs, which signaled limitations in the AI system. Furthermore, the findings showed that example-based explanations supported more appropriate reliance on AI compared to feature-based explanations, which often led to overreliance. For the context of evaluating digital inclusiveness of agri-food digital tools with LLMs, these insights emphasize that replacing or complementing human decisions with AI-driven evaluation must carefully account for explanation design and decision-maker intuition. In agricultural contexts, where inclusiveness assessments involve nuanced socio economic and cultural considerations, ensuring that AI systems promote appropriate reliance rather than blind acceptance is essential for generating trustworthy and equitable evaluations.

## 4. Methodology

This study applied an AI-based rapid assessment to compare LLM-generated with human expert assessments of digital agricultural tools under the MDII framework Four LLMs (GPT-5, GPT-4o, Grok, and Gemini) were tested using a standardized prompt structure derived from MDII dimensions, subdimensions, and indicators to ensure consistency and comparability across tools and models. Nine digital tools were reviewed, spanning applications in irrigation management, crop insurance, and market access. These included platforms for matching farmers with relevant service providers, computing irrigation performance indicators using remote sensing data, supporting the design of crop insurance schemes, and integrating weather and soil information to optimize irrigation advice. Additional platforms focused on assessing water productivity in wheat systems, monitoring irrigation scheme performance and estimating water use through precipitation and evapotranspiration comparisons (see Table 1).

Each tool was evaluated through structured AI prompts reflecting MDII's inclusiveness criteria. Identical contextual information was provided to all models, including tool descriptions, innovator responses, and validated MDII indicators, ensuring that any scoring variation reflected model reasoning rather than input discrepancies.

*Table 1: Descriptions of Tools by ID and Purpose.*

| Tool ID | Purpose |
|---|---|
| **INT_T02** | Matches farmers with appropriate service providers using collected agricultural/farmer data. |
| **WP_T01** | Compute agricultural water use, crop water requirements, and irrigation performance indicators. |
| **WP_T02** | Supports insurance design by estimating yield impacts of droughts/floods using WaPOR data. |
| **WP_T03** | Assesses yield fluctuations to support decisions in irrigated and rainfed wheat systems. |
| **WP_T04** | Assesses irrigation scheme performance using remote sensing and field data. |
| **WP_T05** | Provides water productivity indicators for the Jordan Valley using integrated datasets. |
| **WP_T06** | Combines weather, transpiration, and soil moisture data to optimize irrigation advice. |
| **WP_T07** | Calculates water use by comparing evapotranspiration and precipitation over time. |
| **WP_T09** | Monitors and forecasts drought to government agencies and rural communities to help information in putting in place mitigation measures. |

To ensure methodological reproducibility, all prompts were developed using a standardized template reflecting MDII's evaluation structure. Full prompt templates and examples are included in Annex A. Each prompt consisted of (i) a domain-specific system message defining the expert persona and evaluation criteria, (ii) a set of contextual tool descriptions and indicator-level information, and (iii) an evaluative instruction requesting a score (0–5) and a justification. All evaluations followed a zero-shot prompting approach, with no fine-tuning or few-shot examples provided.

Scoring was performed directly by the LLMs through structured natural language output. Each response included a numeric score and a paragraph-length justification for each subdimension. These outputs were automatically parsed and stored for comparison. No post-processing or manual correction was applied to LLM-generated scores or justifications. The intent was to assess models' ability to autonomously interpret and apply inclusiveness criteria in a controlled evaluation setting.

By isolating the scoring logic within the prompt design and standardizing the context provided, we ensured that model variation could be attributed to internal reasoning pathways rather than external data inconsistencies or prompt ambiguity.

### 4.1 AI-based Rapid Assessment Workflow

The dataset used in this study was drawn from the centralized MDII evaluation database, developed and maintained by the MDII team. This database compiles structured responses from multiple user groups through a standardized MDII online survey process. Data collection combined several methods. Participants completed role-specific survey forms aligned with their functional area, ensuring that responses reflected both technical and experiential perspectives on tool inclusiveness. These data provided the contextual foundation for the AI rapid assessment, serving as the reference corpus against which LLM evaluations were performed (see Figure 2).

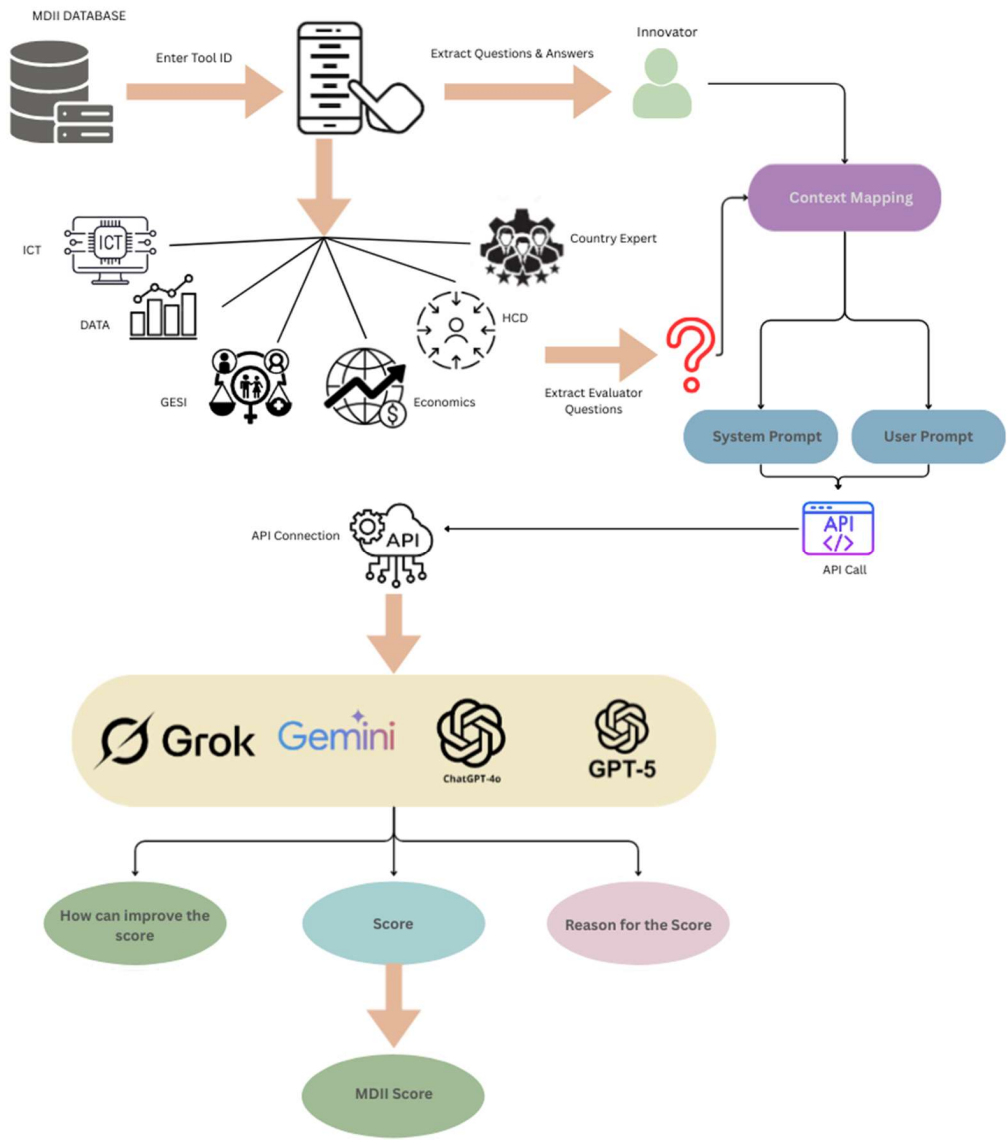

*Figure 3: AI Rapid Assessment Workflow. Source: Authors.*

## 4.2 Context Mapping as a Requirement

To support evaluators in generating consistent and well-informed assessments, a structured context mapping mechanism was integrated into the evaluation workflow. This mechanism provided systematic access to key information that evaluators were required to reference when answering assessment questions. For each digital tool or innovation, evaluators were guided to consider three complementary levels of context. First, a *Tool Summary* offered a high-level overview prepared by the innovator, including the tool's purpose, target users, operating environment, and core functionalities. This summary served as the main reference point for understanding the overall scope and intended outcomes of innovation. Second, *Subdimension Contextual Information* outlined how the innovation related to a specific subdimension of digital inclusiveness (e.g., solution effectiveness, problem relevance, digital accessibility, and long-term loss), ensuring that evaluators interpreted responses through the appropriate analytical lens. Third, where available, *Indicator-Level Context* provided granular clarifications at the indicator level, offering precise reference points against which tools could be assessed. This layered and standardized approach reduced cognitive load for evaluators and enhanced both comparability and internal consistency across assessments. Furthermore, it ensured that both human- and AI-based evaluations were grounded in the same conceptual foundational and interpretive frame regarding each tool's relevance to the MDII framework. For AI-based assessments, this contextual information was formally encoded and embedded into prompts to replicate and scale the evaluative considerations applied by human assessors (see Figure 3).

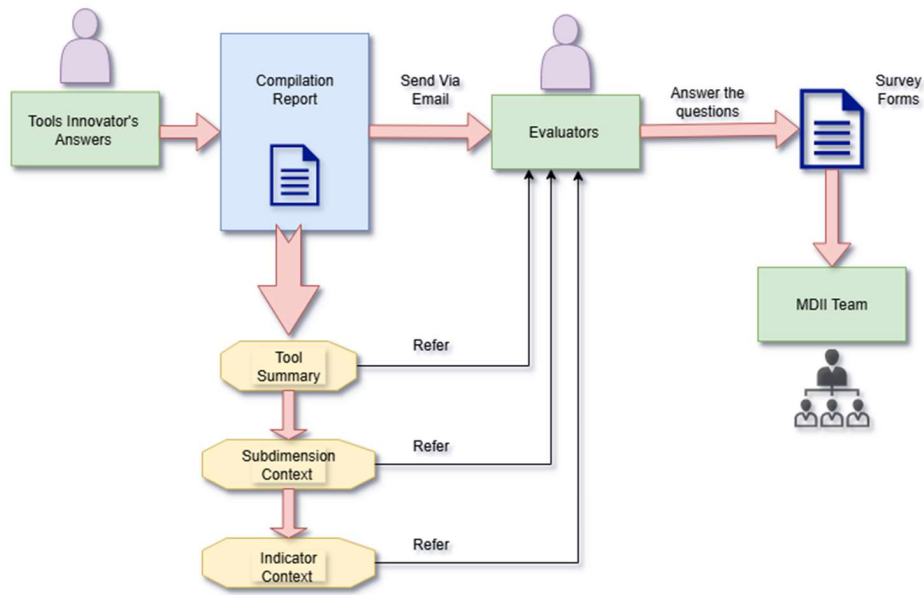

*Figure 4: Context Mapping and data collection process from Evaluators. Source: Authors.*

### 4.3 Experimental Setup

A controlled experimental framework was established to compare the performance of human experts and AI systems in assessing the inclusiveness of digital agri-food tools. The framework was designed to ensure consistency, scalability, and reliability, while accounting for computational and temporal constraints. The experiment involved comparing the outputs of four LLMs: Grok (xAI, n.d.), Gemini (gemini-2.5-flash-lite, 2025), GPT- 4o (OpenAI, Hello GPT-4o, 2024), and GPT-5 (OpenAI, Introducing GPT-5, 2025). For Grom, Gemini, and GPT-4o, assessments were conducted under three temperature settings (0.3, 0.7, and 0.9) to assess how stochastic parameter variation influenced scoring stability.

In contrast, GPT-5 being a reasoning-focused model, does not employ temperature as a configurable parameter. This allowed direct comparison of GPT-5's reasoning stability against parameter variation of temperature-sensitive models. To maintain uniform computational conditions, all models were subject to standardized resource constraints: a limit of 20 requests per minute, a 2.5-second delay between API calls to manage server load and maintain response consistency, and up to five automatic retries for failed requests. A top-p value of 0.9 was applied to constrain token sampling, and a 20,000-token cap ensured full processing of detailed tool descriptions. Both human evaluators and AI systems analyzed identical datasets derived from MDII assessments of digital agrisystem innovations. This setup enabled a controlled, one-to-one comparison of model and human performance, highlighting the relative strengths, weaknesses, and stability of Grok, Gemini, GPT-4o, and GPT-5 in assessing digital inclusiveness.

### 4.4 Generative AI Configuration

GenAI, particularly LLMs, offers new opportunities for applying structured evaluation methods to complex systems such as digital agrisystem tools. In this study, LLMs were configured to generate both quantitative inclusiveness scores (0–5) and qualitative justifications across six expert domains: Information and Communication Technology (ICT), Data, Gender Equality and Social Inclusion (GESI), Economics, Human-Centered Design (HCD), and Country Expertise. Each domain was represented by a specialized expert persona embedded within the system prompts, defining the AI as a senior professional with more than 10 years of experience, relevant certifications, and institutional affiliations. System prompts were tailored to domain specific priorities, instructing the LLMs to critically evaluate digital tools according to global standards and MDII principles. For example, the ICT persona emphasized technological effectiveness and scalability, while the GESI persona focused on equity and social responsiveness. All models received identical contextual inputs from a PostgreSQL database containing validated innovator responses and indicator-level metadata. This ensured a consistent informational baseline across evaluations. By combining domain-specific personas with structured prompts, the LLMs produced context-aware, criterion-based evaluations aligned with MDII principles. This human AI synergy facilitated a more comprehensive analysis of digital inclusiveness, highlighting both strengths and gaps in agrifood systems, and ultimately informing the development of more equitable digital tools.

## Results and Discussion

### 4.1 Performance across stochastic LLMs

Initial analyses compared the performance of Grok, Gemini, and GPT-4o across three temperature settings (0.3, 0.7, and 0.9). Each model generated MDII scores for the same set of digital tools, which were then compared with human expert evaluations to quantify alignment. Across temperature settings, Grok demonstrated the lowest mean absolute error (MAE) and the strongest alignment with human assessments at temperature = 0.9, whereas Gemini and GPT-4o displayed moderate variability[1]. As depicted in Table 2, mean absolute error (MAE) values confirm that prediction error differences across temperatures were modest but systematic, indicating sensitivity to stochastic sampling rather than data content. MAE values increased slightly at intermediate temperatures, reflecting the expected impact of stochastic variation, before improving again at higher temperatures. Across temperatures, Grok consistently achieved the lowest MAE, indicating strong alignment with human evaluations, while Gemini performed moderately well and GPT-4o showed the greatest variability, particularly at higher temperatures. These results confirm that AI systems can approximate human assessment of digital inclusiveness, though model choice and temperature settings influence score variance and alignment consistency across tools.

*Table 2: Mean Absolute Error (MAE) per Model and Temperature.*

| Tool | Temp = 0.3 | | | Temp = 0.7 | | | Temp = 0.9 | | |
|---|---|---|---|---|---|---|---|---|---|
| | Grok | Gemini | GPT-4o | Grok | Gemini | GPT-4o | Grok | Gemini | GPT-4o |
| **WP_T01** | 4% | 4% | 6% | 6% | 6% | 10% | 1% | 4% | 6% |
| **INT_T02** | 14% | 18% | 13% | 16% | 17% | 16% | 7% | 20% | 15% |
| **WP_T02** | 2% | 4% | 11% | 2% | 4% | 15% | 1% | 5% | 10% |
| **WP_T03** | 2% | 6% | 18% | 10% | 8% | 13% | 5% | 6% | 19% |
| **WP_T04** | 1% | 1% | 0% | 2% | 3% | 3% | -2% | 2% | 4% |
| **WP_T05** | 7% | 6% | 14% | 7% | 7% | 13% | 2% | 7% | 13% |
| **WP_T06** | 7% | 13% | 12% | 6% | 13% | 13% | 6% | 14% | 12% |
| **WP_T07** | 11% | 13% | 12% | 6% | 13% | 13% | 6% | 11% | 11% |
| **WP_T09** | 5% | 10% | 8% | 6% | 10% | 11% | 5% | 10% | 12% |
| *Average* | *6%* | *8%* | *10%* | *7%* | *9%* | *12%* | *3%* | *9%* | *11%* |

The heatmap (table 3) summarizes the deviation of AI-generated scores from human scores across tools and temperatures. Greener cells indicate smaller deviations (stronger alignment), while redder shades represent larger divergences. Each cell represents the difference in percentage points between AI and human scores, averaged across indicators per tool. This highlights that Grok's performance improves at higher temperature settings (though not strictly linearly), while Gemini remains relatively stable and GPT-4o shows less consistent alignment.

*Table 3: Deviation of AI-generated MDII scores from human expert scores across temperature settings (0.3, 0.7, 0.9).*

| Tool | | Temp = 0.3 | Temp = 0.7 | Temp = 0.9 |
|---|---|---|---|---|

---

[1] While other performance metrics could have been applied, MAE was selected for its interpretability and suitability for comparing deviations across tools.

| | Human Scores | Grok | Gemini | GPT-4o | Grok | Gemini | GPT-4o | Grok | Gemini | GPT-4o |
|---|---|---|---|---|---|---|---|---|---|---|
| **WP_T01** | 39% | -4% | -4% | -6% | -6% | -6% | -10% | -1% | -4% | -6% |
| **INT_T02** | 63% | -14% | -18% | -13% | -17% | -16% | -17% | -7% | -20% | -15% |
| **WP_T02** | 43% | -2% | -4% | -11% | -2% | -4% | -15% | -1% | -5% | -10% |
| **WP_T03** | 49% | -2% | -6% | -18% | -10% | -8% | -13% | -5% | -6% | -19% |
| **WP_T04** | 42% | 0% | -1% | 0% | -2% | -2% | -3% | 2% | -2% | -4% |
| **WP_T05** | 52% | -7% | -6% | -14% | -7% | -7% | -13% | -2% | -7% | -13% |
| **WP_T06** | 56% | -7% | -13% | -12% | -6% | -13% | -13% | -6% | -14% | -12% |
| **WP_T07** | 47% | -11% | -13% | -12% | -6% | -13% | -13% | -6% | -11% | -11% |
| **WP_T09** | 60% | -5% | -10% | -8% | -6% | -10% | -11% | -5% | -10% | -12% |

*Color scale: green = low deviation (high alignment), red = high deviation (low alignment).*

Overall, these results identify Grok at 0.9 as the most accurate and reliable configuration and establish it as the benchmark LLM for subsequent comparisons with GPT-5, which operates deterministically without temperature adjustments.

### 4.1.1 Comparative alignment with human evaluations

Correlations between human MDII scores and AI evaluations were calculated across three temperature settings for each LLM. As shown in Table 3, Grok demonstrated the strongest overall alignment with human evaluations ($M = .889$, $SD = .052$), followed closely by GPT-4o ($M = .843$, $SD = .048$). In contrast, Gemini exhibited weaker correspondence with human scores ($M = .792$, $SD = .050$). Across models, correlations remained consistently high ($r > .76$), indicating that all three LLMs captured similar relative scoring patterns to a substantial degree. In terms of stability, GPT-4o produced the lowest variability across temperatures, suggesting comparatively robust performance under stochastic sampling. Grok, while achieving the highest mean correlation, showed slightly greater variability than GPT-4o, whereas Gemini displayed the most pronounced fluctuations, indicating higher sensitivity to parameter changes.

*Table 4: Pearson Correlations Between Human MDII Scores and LLM Predictions Across Temperature Settings.*

| LLM | Temp = 0.3 | Temp = 0.7 | Temp = 0.9 | Average Correlation | Std Dev |
|---|---|---|---|---|---|
| Grok | 0.853033 | 0.86664 | 0.9488 | 0.889491 | 0.051812 |
| Gemini | 0.763124 | 0.849584 | 0.761891 | 0.791533 | 0.050277 |
| GPT-4o | 0.793518 | 0.8901 | 0.844417 | 0.842678 | 0.048314 |

These results hint that the evaluated LLMs can approximate human assessment patterns with strong linear association, achieving correlation coefficients above 0.76 across temperature settings. Grok maintained the strongest overall correspondence with human evaluations, while GPT-4o demonstrated superior stability under stochastic variation. Taken together, these findings suggest that both models capture underlying decision logic comparable to expert reasoning, even if through different strengths, Grok in low MAE alignment and GPT-4o in correlation consistency across settings. This complementary performance establishes a robust empirical basis for benchmarking GPT-5, which introduces deterministic reasoning in place of temperature-driven stochasticity.

## 4.2 GPT-5 reasoning stability and context awareness

Building upon the comparative results from Grok, Gemini, and GPT-4o, a final experiment was conducted with GPT-5 o test whether a model operating under deterministic settings (i.e., without exposed sampling parameters) could match or exceed the performance of temperature-optimized stochastic configurations achieved by previous models. This transition from temperature-tuned sampling to reasoning under fixed generation settings allowed the study to isolate whether consistency could be achieved through contextual understanding rather than parameter control. GPT-5 was accessed through an interface that did not expose temperature or top-p parameters. As such, all outputs reflect the model's default deterministic-like behaviour under fixed sampling conditions, without stochastic variation. This configuration differs from other models like Grok, which allow explicit temperature tuning.

Using the same dataset and MDII framework domains, GPT-5 generated more coherent, less dispersed scores across tools, producing more consistent alignment with human evaluations, with reduced variance across tools. As shown in Figure 4, GPT-5's results converge closely with expert scores with smaller cross-tool fluctuations. This suggests that GPT-5, under these generation settings, may be leveraging its internal representation capabilities to reflect the latent structure of MDII scoring, understanding inter-dimensional relationships (e.g., between usability, governance, and equity) instead of relying on probabilistic sampling.

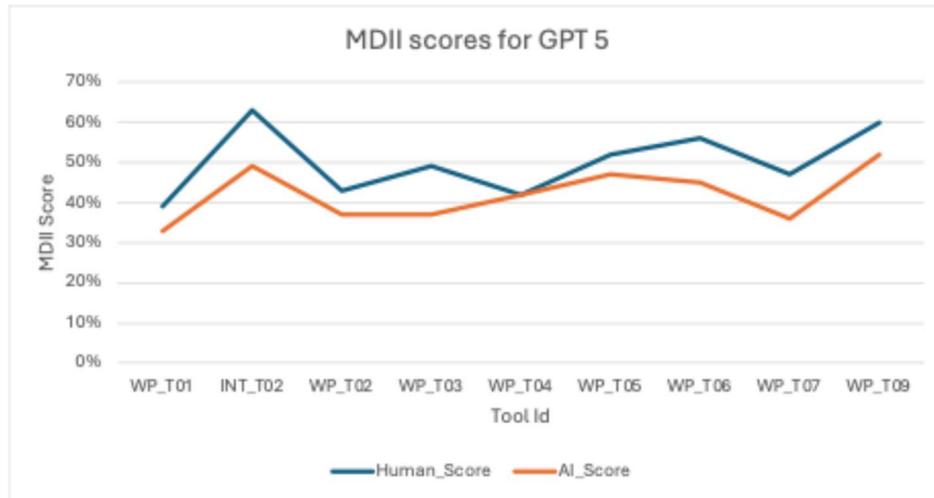

*Figure 5: MDII Scores generated by GPT-5 vs. Human-based MDII scores for each tool. Source: Authors.*

Furthermore, as shown in Figure 5, the human-evaluated scores (blue) serve as the baseline for assessing model performance. The Grok (0.9) curve (orange) follows this reference most closely, reproducing the general shape of human evaluations but with sharper oscillations across tools. In contrast, GPT-5 (green) generates consistently lower yet smoother scores, reflecting its reasoning-based inference over sampling-based variability. This pattern illustrates two complementary behaviours: Grok's temperature-tuned configuration captures local variability, amplifying peaks such as INT_T02 and WP_T09, whereas GPT-5 compresses these fluctuations, producing a more uniform response profile. The flatter green trajectory suggests that GPT-5 prioritises contextual coherence over stochastic exploration, yielding higher internal stability but reduced sensitivity to subtle score differentials.

Alignment with the human baseline is strongest for WP_T04, WP_T09, and WP_T01–02, where input data were more complete and contextually rich. Divergences widen for WP_T03–T07, largely reflecting missing or ambiguous innovator responses. Taken together, these results confirm that GPT-5 human scoring logic with lower inter-tool score variance and more deterministic output behaviour, while Grok approximates human variability more closely but with higher stochastic volatility. Both behaviours highlight that data completeness remains the principal determinant of human-model alignment.

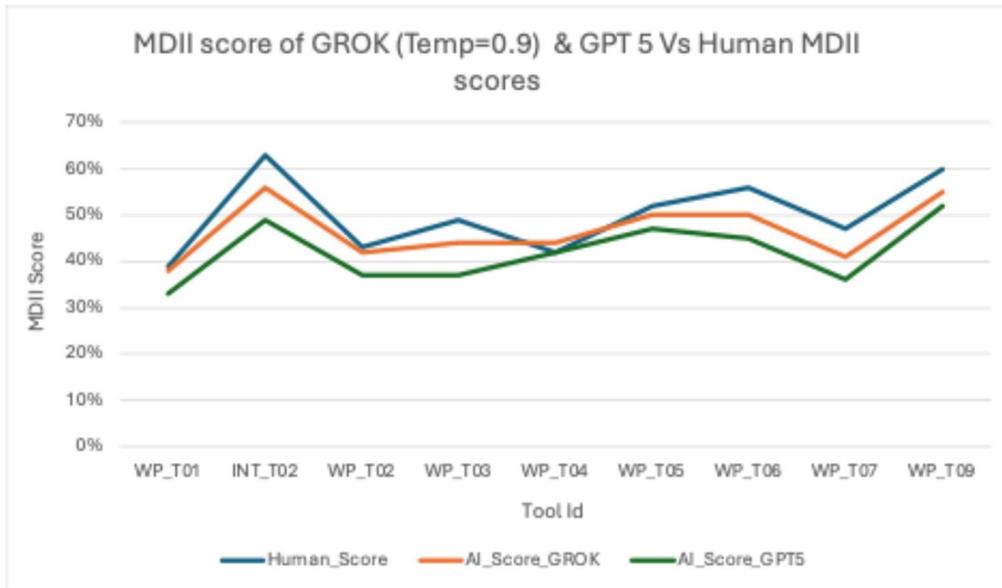

*Figure 6: MDII Scores generated by GPT-5 and GROK vs. Human-based MDII scores for each tool. Source: Authors.*

Overall, the comparative results between GPT-5 and Grok (temp = 0.9) demonstrate that deterministic-like reasoning can reach functional equivalence with optimally tuned stochastic models in replicating expert evaluations of digital inclusiveness. The smoother and more conservative scoring trajectory observed in GPT-5 suggests that contextual reasoning, rather than parameter adjustment, now drives stability in scoring behaviour. Despite their architectural and configuration differences, both models demonstrate that reliability is primarily determined by the completeness, structure, and contextual coherence of the tool-specific input data, even under identical evaluation conditions. If the data source is consistent across all tools, the variation observed does not result from differences in data origin but from differences in data completeness and contextual richness at the tool level. Although all models use the same evaluation framework and indicators, each tool (for example, WP_T01 or WP_T03) provides its own set of innovator responses and contextual details. Some tools offer more complete and coherent information, with fewer missing answers and clearer descriptions, while others are less detailed or more fragmented. These differences in information quality explain why alignment between human and AI evaluations is stronger for some tools, even when the experimental setup remains identical.

### 4.3 Influence of Data Completeness on Evaluation Reliability

To better understand divergences in alignment across tools, we analyzed the number of unanswered questions in the original innovator submissions. As shown in Table 5, tools such as WP_T03 (28), WP_T05 (29), and INT_T02 (25) had the highest number of missing responses. These tools also showed the largest discrepancies between AI-generated and human-evaluated MDII scores.

In contrast, tools like WP_T01, WP_T02, and WP_T04 had significantly fewer unanswered items (8–16) and demonstrated much closer score alignment across models, especially for GPT-5 and Grok (see Figure 6). These findings suggest that input completeness is a critical driver of score stability, affecting model outputs.

*Table 5: Number of Questions Unanswered by Innovators Across Tools.*

| Tool ID | WP_T01 | WP_T02 | WP_T03 | WP_T04 | WP_T05 | WP_T06 | WP_T07 | WP_09 | INT_02 |
|---|---|---|---|---|---|---|---|---|---|
| Unanswered Questions | 8 | 16 | 28 | 14 | 29 | 23 | 22 | 12 | 25 |

Dimensional score comparisons further reinforce this pattern. For example, as seen previously on Figure 6, it shows that WP_T04, a tool with high response completeness, achieved the strongest alignment across models and human evaluations. WP_T03, WP_T05, and WP_T06, which had more missing or ambiguous data, showed the weakest alignment regardless of the model used. This consistency across tools and models suggests that score reliability is not solely a function of model choice or configuration, but also depends heavily on the clarity, depth, and completeness of the qualitative data provided. Variation in data collection methods may also contribute to these differences. Tools like WP_T04 involved guided virtual interviews using Microsoft Teams, whereas tools like WP_T05 relied on standalone online form submissions. This discrepancy in data collection modality may help explain variations in response depth and, consequently, in the alignment between AI-generated and human-evaluated scores. Taken together, these results indicate that the success of LLM-based evaluation pipelines depends as much on the structure and completeness of input data as on the sophistication of the models themselves. Model performance in this context should be understood as data-conditioned reliability rather than intrinsic capability.

## 5. Conclusions and Implications

The comparative analysis suggests that LLMs can partly replicate human evaluation logic within the MDII framework, although they appear to do so through different reasoning mechanisms. Stochastic models such as Grok, Gemini, and GPT-4o rely on parameter tuning to manage variation, while GPT-5 seems to achieve similar alignment through contextual reasoning when run under deterministic-like settings (i.e., non-configurable sampling settings). These results are preliminary but point to a possible shift from parameter-based optimisation to reasoning-based consistency. Across models, the strongest alignment with human evaluations occurs in tools with more complete and contextually detailed data. This confirms that LLMs can approximate human evaluators, but only when provided with complete and context-rich input. Since all models were tested under identical conditions, the observed differences likely stem from variation in data quality rather than model bias. When information is limited or unclear, even reasoning-oriented models show inconsistency, similar to human evaluators working with incomplete evidence. The results also indicate that reasoning-based architectures may begin to recognise relationships among MDII dimensions, such as how usability relates to governance or how social inclusion influences perceived benefit. This emerging ability could be useful for improving the structure of future AI-supported evaluations. In practice, AI-enabled assessments could complement human evaluations by providing faster and more reproducible results, especially in early-stage or resource-limited contexts. However, the ongoing need for robust data curation pipelines and context-rich documentation to preserve interpretive fidelity. Integrating LLMs into MDII workflows should therefore focus on data quality assurance and methodological transparency rather than on model substitution or overfitting.

At the institutional level, the results suggest that temperature-tuned models such as Grok (temperature = 0.9) currently offer the strongest alignment with human evaluations, particularly when input data is complete and contextually coherent. In parallel, reasoning-based models such as GPT-5 appear to deliver greater scoring stability and smoother transitions across MDII dimensions. These findings do not indicate categorical superiority of one architecture over another, but rather highlight distinct performance profiles: Grok demonstrates higher alignment when optimally tuned, while GPT-5 shows greater consistency and interpretability across cases. Future implementations should consider model selection based on the specific priorities of each evaluation context: whether the emphasis is on alignment (Grok), reproducibility (GPT-5), or broader generalisability across tools and settings.

## 6. Limitations

This study presents several limitations that should be considered when interpreting its findings. First, the evaluation process depends heavily on prompt design, as current LLMs cannot access external files, URLs, or embedded documents. All relevant context must therefore be encoded within the prompt itself, which places practical constraints on the volume and nuance of information that can be conveyed. This can result in variable performance depending on the clarity and completeness of the input, particularly when prompts must balance brevity with interpretive precision. Second, while previous generative models allowed parameter tuning (e.g., temperature) to modulate response variability, GPT-5 operates under a more deterministic regime, reducing researchers' control over response diversity. This may influence the extent to which exploratory or edge-case scenarios can be reliably tested. Third, the limited number of tools (n=9) and domain-specific inputs restricts generalisability. Although all models were run under consistent conditions, the relatively small sample size increases the risk that outputs may be overly influenced by specific details or wording in the tool descriptions, rather than general patterns.

Moreover, the limited dataset introduces the potential for Type I and Type II errors in model predictions. A Type I error (false positive) would occur if an AI model incorrectly scores a tool as inclusive despite lacking sufficient evidence, potentially leading to unwarranted trust or prioritization. Conversely, a Type II error (false negative) could result in the dismissal or de-prioritization of a tool that is, in fact, aligned with inclusiveness principles, simply due to incomplete input. While the controlled evaluation setup minimizes these risks, the narrow sample size and variability in tool documentation limit the generalisability of these safeguards. Lastly, while GPT-5 was accessed under fixed-generation conditions via API (without temperature or top-p exposure), this does not imply architectural determinism. The term 'deterministic settings' is used here to reflect sampling constraints rather than model design.

## 7. Ethical Considerations and Risk Assessment

The use of LLMs in MDII evaluations introduces new methodological possibilities but also ethical concerns. While this study focuses on technical feasibility, its findings also point to limitations and risks that could affect both the reliability and legitimacy of AI-assisted assessments. The most immediate concern is opacity in model reasoning. Although all models replicated structured evaluation logic with high fidelity, their internal decision-making processes remain non-transparent. Even in deterministic-like configurations, the pathways through which justifications are derived cannot be fully audited or explained, raises the risk of perceived objectivity (i.e., where users trust the output because it mimics expert patterns, without understanding how conclusions are formed). In practice, this suggests that model accuracy alone is not sufficient. The existence of traceability and interpretability must be designed into any future deployment to avoid institutional overreliance on AI-generated outputs[2]. This opacity raises the risk not only of misinterpretation but also of systematic Type I and II errors in institutional decision-making. Without visibility into how judgments are formed, institutions may unknowingly act on flawed outputs, either by accepting tools that do not meet inclusion standards (Type I) or rejecting those that do (Type II). In contexts where evaluations influence funding, partnerships, or reputational visibility, such errors could have cascading consequences.

Secondly, the results highlight how bias can emerge not from the model itself, but from input asymmetries across tools. All models were exposed to the same framework and indicators, but the underlying innovator datasets varied in completeness and contextual richness. Tools with clearer, more complete documentation produced more stable outputs across both human and AI evaluations. This suggests that inclusiveness assessments are highly sensitive to input quality and that any AI system trained or fine-tuned on real-world data can amplify structural inequalities already embedded in the evaluation ecosystem. These findings reinforce the notion that LLMs are not neutral instruments because they operationalise whatever data they are given. Therefore curation practices to avoid underrepresenting tools developed in low-data contexts should be present at all times. Moreover, the (perceptual) efficiency of AI scoring introduces new forms of automation bias. When LLMs generate seemingly well-reasoned scores at scale, there is a risk that institutions might bypass necessary human review processes. This risk is not hypothetical: in high-volume or resource-constrained settings, the incentives to use AI as a shortcut for evaluation will grow. Yet as shown in this study, even small gaps in contextual clarity can produce interpretive divergence between models and human experts. This reinforces the need for governance protocols that define how AI outputs are reviewed, validated, or contested, particularly when assessments influence funding, programming, or reputational outcomes. Privacy and data rights also need to be considered. While this experiment used anonymised responses, operational use of LLMs would require formal policies on data reuse, informed consent, and secondary processing. As LLMs scale, so do questions of data ownership, especially when innovator responses are used to generate scores or justifications beyond the original intent. Alignment with FAIR and TRUST principles, and the development of privacy-preserving architectures, will be essential for maintaining stakeholder trust.

---

[2] Such design principles align with broader normative frameworks, including the OECD AI Principles and UNESCO's Recommendation on the Ethics of Artificial Intelligence, which emphasise transparency, explainability, and proportional human oversight in AI-assisted decision-making.

Finally, the question of accountability remains unresolved. In this study, the authors maintained full oversight of scoring logic and data interpretation. But in operational contexts, responsibility for errors or unintended consequences may become more diffuse. Institutions adopting AI-supported assessments will need to delineate where accountability lies: whether with the model, prompt design, underlying dataset or human decision-maker. Without such clarity, AI adoption could blur responsibility rather than distribute it. Taken together, these findings suggest that technical advances in model performance must be matched by advances in ethical oversight. If LLMs are to support MDII evaluations in practice, their integration must be accompanied by interpretability safeguards, bias mitigation strategies, and human-in-the-loop validation.

## 8. Future Work

Future work should examine the generalisability of AI-assisted evaluations across a broader set of digital tools and implementation contexts. While this study provides an initial proof of concept, the limited sample size restricts the extent to which findings can be extrapolated. Subsequent research should include tools developed in different languages, institutional settings, and data environments to assess whether current alignment patterns persist. Methodological extensions should also explore the influence of prompt design, input encoding, and evaluation granularity on model behaviour. Additionally, studies should test human-in-the-loop validation workflows to understand how expert oversight interacts with model outputs and whether interpretability interventions affect evaluation reliability.

Finally, future efforts should address the operational requirements for scaling AI-based assessments, including documentation standards, data governance protocols, and ethical review mechanisms that ensure transparency and accountability in automated evaluation processes. Future studies should also investigate how different model architectures, such as parameter tuned models like Grok and reasoning-based models like GPT- 5, perform under varying evaluation demands to inform appropriate model selection.